\documentclass[10pt,fleqn]{article}
\usepackage{amsmath,graphicx,latexsym}

\begin{document}

\title{Integrable mappings derived from the $\Delta \Delta RsG$ equation}

\author{Apostolos Iatrou \thanks{{\copyright\mbox{ }Apostolos Iatrou}}
\thanks{{\sf email: apostolosiatrou@hotmail.com}}}

\date{}

\maketitle


\begin{abstract}

In this paper we consider the integrability of the mappings
derived from the double discrete related sine-Gordon $(\Delta
\Delta RsG)$ equation, which we recently introduced in the paper
{\em Higher dimensional integrable mappings}, under an appropriate
periodicity condition.

\end{abstract}


\section{Introduction}

In this paper we consider the double discrete related sine-Gordon
$(\Delta \Delta RsG)$ equation we recently introduced in
\cite{ai}, i.e.
\begin{equation}\label{rsg}
  V_{l+1,m+1} = -V_{l,m}+\frac{b-a(V_{l+1,m}+ V_{l,m+1})}
  {c(V_{l+1,m}+ V_{l,m+1})+a}.
\end{equation}
We investigate the integrability of the mappings derived under the
periodicity condition $V_{l+\zeta_2,m-\zeta_1}=V_{l,m}$, with
$\zeta_1$ and $\zeta_2$ being relatively prime.\footnote{For a
detailed discussion of mappings derived from partial difference
equations under such a periodicity condition see \cite{qcpn}.}
Under this periodicity condition, we derive the
$(\zeta_1+\zeta_2)$ dimensional map
\begin{equation}\label{map}
  V_{0}'=V_{1}\, , \; \; \dots \, , \; \;
  V_{\zeta_1+\zeta_2-2}'=V_{\zeta_1+\zeta_2-1} \, , \; \;
  V_{\zeta_1+\zeta_2-1}'=-V_0+\frac{b-a(V_{\zeta_1}+ V_{\zeta_2})}
  {c(V_{\zeta_1}+ V_{\zeta_2})+a},
\end{equation}
where $V_{l,m}=V_n$ and $n=\zeta_1 l+\zeta_2 m$. When the map
(\ref{map}) is odd dimensional, it can be reduced to an even
dimensional map under the reduction $v_i = V_i +V_{i+1}$.

The plan  of this paper is as follows : in section 2 we show that
the mapping (\ref{map}), when $\zeta_1$ odd and $\zeta_2$ even or
$\zeta_1$ even and $\zeta_2$ odd, is related to the mapping
derived from the double discrete (alternative) Korteweg-de Vries
(a $\Delta \Delta KdV$) equation (given in \cite{ai}) when
$\zeta_1$ odd and $\zeta_2$ or $\zeta_1$ even and $\zeta_2$ odd,
respectively. In section 3 we show that in the special case
$\zeta_1 = 1$ and $\zeta_2$ even, the mapping (\ref{map}) is a
particular case of the {\em hierarchy of integrable asymmetric
mappings}\footnote{We call a mapping, which possesses at least one
cyclic invariant $n$-quadratic integral, {\em symmetric} if the
cyclic invariant integral is invariant under any permutation of
the variables, see \cite{ai}. We call a mapping {\em asymmetric}
if the $n$-quadratic integral is not invariant under any
permutation of the variables.} we introduced in \cite{ai}.

\section{Double discrete related
sine-Gordon $(\Delta \Delta RsG)$ equation}

Consider the alternative $\Delta \Delta KdV$ ($a$ $\Delta \Delta
KdV$) equation given in \cite{ai}, i.e.
\begin{equation}\label{kdvmap}
  V_{l+1,m+1}=V_{l,m}+\frac{(\epsilon-\gamma)(V_{l+1,m}-V_{l,m+1})+\xi}
  {\beta(V_{l+1,m}-V_{l,m+1})+\gamma}.
\end{equation}
There are three possible cases to consider using the periodicity
condition $V_{l+\zeta_2,m-\zeta_1}=V_{l,m}$, 1) $\zeta_1$ odd and
$\zeta_2$ even, 2) $\zeta_1$ even and $\zeta_2$ odd, and 3)
$\zeta_1$ odd and $\zeta_2$ odd.

Case 1) When $\gamma=\epsilon=0$ together with
\begin{equation}\label{tran1}
  V_{l,m} \rightarrow \left\{ \begin{array}{c}
    V_{l,m} \, , \; \; \,  \; \; l \,  \; \; \mathrm{even}\\
    -V_{l,m} \, , \; \; \,  \; \; l \,  \; \; \mathrm{odd} \
  \end{array}
  \right.
\end{equation}
we obtain
\begin{equation}\label{kdvmap1}
  V_{l+1,m+1}=-V_{l,m}+\frac{\xi}
  {\beta(V_{l+1,m}+V_{l,m+1})}.
\end{equation}

Case 2) When $\gamma=\epsilon=0$ and
\begin{equation}\label{tran2}
  V_{l,m} \rightarrow \left\{ \begin{array}{c}
    V_{l,m} \, , \; \; \,  \; \; m \,  \; \; \mathrm{even}\\
    -V_{l,m} \, , \; \; \,  \; \; m \,  \; \; \mathrm{odd} \
  \end{array}
  \right.
\end{equation}
we obtain
\begin{equation}\label{kdvmap2}
  V_{l+1,m+1}=-V_{l,m}-\frac{\xi}
  {\beta(V_{l+1,m}+V_{l,m+1})}.
\end{equation}
In cases 1) and 2) the periodicity condition remains invariant
under the transformation (\ref{tran1}) and (\ref{tran2}) and as a
result the mapping derived from (\ref{kdvmap}) (with
$\gamma=\epsilon=0$) is the same as that derived from
(\ref{kdvmap1}) or (\ref{kdvmap2}). In case 3), however, the
periodicity condition does not remain invariant under the
transformation (\ref{tran1}) and (\ref{tran2}) and as a result the
mapping derived is different. We will not consider case 3) in this
paper.

The mapping derived from (\ref{kdvmap1}) or (\ref{kdvmap2}) under
the periodicity condition $V_{l+\zeta_2,m-\zeta_1}=V_{l,m}$ is
\begin{equation}\label{map1}
  V_{0}'=V_{1}\, , \; \; \dots \, , \; \;
  V_{\zeta_1+\zeta_2-2}'=V_{\zeta_1+\zeta_2-1} \, , \; \;
  V_{\zeta_1+\zeta_2-1}'=-V_0 \pm \frac{\xi}
  {\beta(V_{\zeta_1}+ V_{\zeta_2})},
\end{equation}
with $+$ sign for case 1) and $-$ sign for case 2). Using the
translations $V_i \rightarrow V_i + a/2\beta$ and $V'_i
\rightarrow V'_i + a/2\beta$ and setting $(\beta,\xi)= (c,\pm
(a^2/c + b))$, the mapping (\ref{map1}) becomes the mapping
(\ref{map}).

We now turn to the integrals of the mapping (\ref{map}). We
consider case 1) only, case 2) is similar. Under the periodicity
condition $V_{l+\zeta_2,m-\zeta_1}=V_{l,m}$ and
$\gamma=\epsilon=0$, (\ref{kdvmap}) can be used to derive the
$(\zeta_1+\zeta_2)$ dimensional map
\begin{equation}\label{map2}
  V_{0}'=V_{1}\, , \; \; \dots \, , \; \;
  V_{\zeta_1+\zeta_2-2}'=V_{\zeta_1+\zeta_2-1} \, , \; \;
  V_{\zeta_1+\zeta_2-1}'=V_0+\frac{\xi}
  {\beta(V_{\zeta_1}- V_{\zeta_2})}.
\end{equation}
Under the transformations $V_i \rightarrow (-1)^i V_i$ and $V'_i
\rightarrow (-1)^i V'_i$ the map (\ref{map2}) becomes
\begin{equation}\label{map3}
  V_{0}'=-V_{1}\, , \; V_{1}'=-V_{2} \; \, , \; \; \dots \, , \; \;
  V_{\zeta_1+\zeta_2-2}'=-V_{\zeta_1+\zeta_2-1} \, , \; \;
  V_{\zeta_1+\zeta_2-1}'=V_0-\frac{\xi}
  {\beta(V_{\zeta_1}+ V_{\zeta_2})}.
\end{equation}
We note that the map (\ref{map3}) can be written as
$L_{\ref{map3}} = L \circ L_{\ref{map1}} = L_{\ref{map1}} \circ
L$, where $L$ is defined to be the map with coordinates $V'_i = -
V_i$ (for all $i$) and $L_{\ref{map1}}$ is the map (\ref{map1}).
Using this, we see that $L_{\ref{map3}} \circ L_{\ref{map3}} = L
\circ L_{\ref{map1}} \circ L \circ L_{\ref{map1}} = L \circ L
\circ L_{\ref{map1}} \circ L_{\ref{map1}} =L_{\ref{map1}} \circ
L_{\ref{map1}}$, as $L$ is an involution, i.e. $L \circ L =
\mathrm{id}$, where $\mathrm{id}$ is the identity map. This shows
that for all integrals $I_i$ of the map $L_{\ref{map3}}$, we have
$I_i(V_0, \dots, V_{\zeta_1+\zeta_2-1}) = I_i(-V'_0, \dots,
-V'_{\zeta_1+\zeta_2-1}) = I_i(V''_0, \dots,
V''_{\zeta_1+\zeta_2-1})$ for the map (\ref{map1}). In fact
$I_i(V_0, \dots, V_{\zeta_1+\zeta_2-1}) = -I_i(V'_0, \dots,
V'_{\zeta_1+\zeta_2-1})$, i.e. the actual integrals of the mapping
(\ref{map1}) are $I^2_i$.

\section{$\zeta_1 = 1$ and $\zeta_2$
even}

In this section we show that in the special case $\zeta_1 = 1$ and
$\zeta_2$ even, the mapping (\ref{map}) is a particular case of
the hierarchy of integrable asymmetric mappings mentioned above.
We illustrate this with the four-dimensional case, i.e.
$\zeta_1=1$ and $\zeta_2=4$.

When $\zeta_1 = 1$ and $\zeta_2=4$, the mapping (\ref{map})
becomes
\begin{equation}\label{rsg14}
  V_{0}'=V_{1}\, , \; \; V_{1}'=V_{2} \, , \; \;
  V_{2}'=V_{3} \, , \; \; V_{3}'=V_{4}\, , \; \;
  V_{4}'=-V_0+\frac{b-a(V_1+ V_4)}
  {c(V_1+ V_4)+a}
\end{equation}
Under the reduction $w = V_0 + V_1$, $x = V_1 + V_2$, $y = V_2 +
V_3$ and $z = V_3 +V_4$, we obtain the four-dimensional mapping
\begin{equation}\label{rsg4}
  w'=x\, , \; \; x'=y \, , \; \;
  y'=z \, , \; \;
  z'=-w+\frac{c(x-y+z)^2+b}{c(x-y+z)+a}.
\end{equation}
Consider the four-dimensional case of the hierarchy of integrable
asymmetric mappings given in \cite{ai}, i.e. the mapping
\begin{eqnarray}\label{L4}
w'&=&-w-\frac{\beta(x+y+z)^2+\epsilon(x+y+z)+\xi_0}{\beta(x+y+z)+\gamma_0}\nonumber
  \\
x'&=&-x-\frac{\beta(w'+y+z)^2+\epsilon(w'+y+z)+\xi_1}{\beta(w'+y+z)+\gamma_1}\nonumber
  \\
y'&=&-y-\frac{\beta(w'+x'+z)^2+\epsilon(w'+x'+z)+\xi_2}{\beta(w'+x'+z)+\gamma_2}\nonumber
  \\
z'&=&-z-\frac{\beta(w'+x'+y')^2+\epsilon(w'+x'+y')+\xi_3}{\beta(w'+x'+y')+\gamma_3},
\end{eqnarray}
Under the substitutions $(w,x,y,z) \rightarrow (w,-x,y,-z)$ and
$(w',x',y',z') \rightarrow (w',-x',y',-z')$ the mapping (\ref{L4})
becomes
\begin{eqnarray}\label{L4s}
w'&=&-w+\frac{\beta(x-y+z)^2-\epsilon(x-y+z)+\xi_0}{\beta(x-y+z)-\gamma_0}\nonumber
  \\
x'&=&-x+\frac{\beta(w'+y-z)^2+\epsilon(w'+y-z)+\xi_1}{\beta(w'+y-z)+\gamma_1}\nonumber
  \\
y'&=&-y+\frac{\beta(-w'+x'+z)^2-\epsilon(-w'+x'+z)+\xi_2}{\beta(-w'+x'+z)-\gamma_2}\nonumber
  \\
z'&=&-z+\frac{\beta(w'-x'+y')^2+\epsilon(w'-x'+y')+\xi_3}{\beta(w'-x'+y')+\gamma_3},
\end{eqnarray}
If we set $\xi_0 = \xi_1 = \xi_2 = \xi_3 = \xi$, $\gamma_0 =
-\gamma_1 = \gamma_2 = -\gamma_3 =-\gamma$ and $\epsilon=0$ the
mapping (\ref{L4s}) can be written as $L_1^4 = L_1 \circ L_1 \circ
L_1 \circ L_1$, where $L_1$ is given by
\begin{equation}\label{L1}
  \tilde{w} = x \, , \; \; \tilde{x} = y \, , \; \; \tilde{y} = z \, , \; \;
  \tilde{z} = -w+\frac{\beta(x-y+z)^2+\xi}{\beta(x-y+z)+\gamma},
\end{equation}
which is equivalent to (\ref{rsg4}). The integrals, $I_1$ and
$I_2$ of the mapping (\ref{L1}), can be obtained from the
coefficients of the different powers of $\lambda$ of the
characteristic equation
\begin{equation}\label{char}
  C(h,\lambda)=\det(\lambda I - L(h)) = 0,
\end{equation}
where $I$ is the identity matrix and $L(h)$ is the $L$ matrix of
the Lax pair ($L,M$) for the mapping (\ref{L4}), see \cite{ai}. We
note that the integrals, $I_1$ and $I_2$, satisfy
$I_i(w,x,y,z)=-I_i(w',x',y',z')$, for $i=1,2$, i.e the actual
integrals for the mapping (\ref{L1}) are $\bar{I}_i = I^2_i$.
\footnote{The symplectic structure for the mapping (\ref{L1}) is
given in \cite{ai}. Using this symplectic structure, one can
easily show that the two functionally independent integrals,
$\bar{I}_1$ and $\bar{I}_2$, are in involution.} The above
discussion also shows that (\ref{L1}) has an asymmetric form, i.e.
(\ref{L4s}), which is equivalent to (\ref{L4}).

This procedure can be applied to any even $\zeta_2$ (with
$\zeta_1=1$). Showing that the mappings obtained from the $(\Delta
\Delta RsG)$, with $\zeta_1=1$ and $\zeta_2$ even, are special
cases of the hierarchy of integrable asymmetric mappings
introduced in \cite{ai}.

\section{Conclusion}

In this paper we have shown that the mappings derived from the
$(\Delta \Delta RsG)$ equation, when $\zeta_1$ odd and $\zeta_2$
even or $\zeta_1$ even and $\zeta_2$ odd, are related to the
mappings derived from the $a$ $\Delta \Delta KdV$ equation, when
$\zeta_1$ odd and $\zeta_2$ or $\zeta_1$ even and $\zeta_2$ odd,
respectively. We have also shown that the mappings derived from
the $(\Delta \Delta RsG)$ equation, when $\zeta_1=1$ and $\zeta_2$
even, are special cases of the hierarchy of integrable asymmetric
mappings given in \cite{ai}. We note that a similar relationship
exists between the asymmetrized mappings\footnote{There are two
minor errors in \cite[appendix C] {ai} : 1) the transformation
reducing the asymmetric ($2m+1$)-dimensional mappings to
($2m$)-dimensional mappings is actually $V_i=v_i v_{i+1}$, and 2)
two lines below the previous error, $v_i=V_i V_{i+1}$.} obtained
from the $a$ $\Delta \Delta MKdV$ and $a$ $\Delta \Delta sG$ (when
$\zeta_1=1$ and $\zeta_2$ even) equations, see \cite[Aappendix
C]{ai}. An open question is whether a transformation exists
relating these two hierarchies? If a transformation relating these
two hierarchies does exist, then the hierarchy of asymmetric
mappings given in \cite[Aappendix C]{ai} is integrable (which we
expect). Another open question is whether the mappings derived
from the $a$ $\Delta \Delta KdV$, $a$ $\Delta \Delta
MKdV$\footnote{See \cite{ai} for the prefix $a$.}, $a$ $\Delta
\Delta sG^5$ and $\Delta \Delta RsG$ equations under an
appropriate periodicity condition are special cases of the
hierarchy of integrable asymmetric mappings given in \cite{ai}.

\section*{Appendix A. Intermediate forms}

In \cite{ai} we showed how to determine intermediate forms (i.e.
where some symmetry in the variables exists) that are integrable
using nonautonomous mappings. In this appendix we show another way
of obtaining such forms, but this time using the hierarchy of
autonomous integrable asymmetric mappings. We illustrate with the
four dimensional case.

Consider (\ref{L4}) when $\xi_2 = \xi_0$, $\xi_3 = \xi_1$,
$\gamma_2 = \gamma_0$ and $\gamma_3 = \gamma_1$, i.e.
\begin{eqnarray}\label{L4sym}
w'&=&-w-\frac{\beta(x+y+z)^2+\epsilon(x+y+z)+\xi_0}{\beta(x+y+z)+\gamma_0}\nonumber
  \\
x'&=&-x-\frac{\beta(w'+y+z)^2+\epsilon(w'+y+z)+\xi_1}{\beta(w'+y+z)+\gamma_1}\nonumber
  \\
y'&=&-y-\frac{\beta(w'+x'+z)^2+\epsilon(w'+x'+z)+\xi_0}{\beta(w'+x'+z)+\gamma_0}\nonumber
  \\
z'&=&-z-\frac{\beta(w'+x'+y')^2+\epsilon(w'+x'+y')+\xi_1}{\beta(w'+x'+y')+\gamma_1}.
\end{eqnarray}
The mapping (\ref{L4sym}) can be written as $L= L_2^2 = L_2 \circ
L_2$, where $L_2$ is given by
\begin{eqnarray}\label{L2}
  \hat{w} &=& y \nonumber \\ \hat{x} &=& z \nonumber \\
  \hat{y} &=& -w-\frac{\beta(x+y+z)^2+\epsilon(x+y+z)+\xi_0}{\beta(x+y+z)+\gamma_0} \nonumber \\
  \hat{z} &=&
  -x-\frac{\beta(y+\hat{y}+z)^2+\epsilon(y+\hat{y}+z)+\xi_1}{\beta(y+\hat{y}+z)+\gamma_1}.
\end{eqnarray}
This mapping is equivalent to the mapping (D.4) of \cite{ai}. In
fact, if we make the substitutions $(w,x,y,z) \rightarrow
(w,y,x,z)$ and $(\hat{w},\hat{x},\hat{y},\hat{z}) \rightarrow
(\hat{w},\hat{y},\hat{x},\hat{z})$ in the mapping (\ref{L2}) we
obtain the mapping (D.4). The integrals of (D.4) can be obtained
in the same way, noting that the integrals of (\ref{L2}) can be
obtained from the Lax pair given in \cite{ai}.

More generally, the intermediate forms can be determined in the
following way : find the possible integers $i$ that divide $n$,
e.g for $n = 16$ we have $i=1,2,4,8,16$. Set the $\gamma_k$'s and
$\xi_k$'s to $\gamma_l$ and $\xi_l$ $(l=0,\dots, i-1)$ in such a
way that we have $i$ of each\footnote{This is achieved by setting
the first $i$ $\gamma_k$'s and $\xi_k$'s $(k = 0, \dots, i-1)$ to
$(\gamma_0,\dots,\gamma_{i-1})$ and $(\xi_0,\dots,\xi_{i-1})$,
respectively, and then repeating the process another $i-1$ times
for the remaining $\gamma_k$'s and $\xi_k$'s, see example above.}
and the resulting asymmetric map can be written as $L= L^{n/i}_i$.
The intermediate form is $L_i$. \footnote{This way of constructing
intermediate forms can also be applied to the asymmetric mappings
given in \cite[appendix C] {ai}.}

The advantages in this method are 1) the mappings obtained need
not be in general complex, whereas the mappings obtained using the
method given in \cite{ai} are, and 2) the intermediate forms are
easily obtained.

\section*{Appendix B. Involutions}

In this appendix we show that all the mappings given in this
paper, and the mappings given in \cite{ai} and \cite{ai3d}, can be
written as a product (composition) of involutions.

First, we note that the cyclic shift, $L_c$, i.e.
\begin{equation}\label{cyclic}
  v_0'=v_1 \, , \; \; \dots \, , \; \; v_{n-1}'=v_n \, , \; \;
  v_n'=v_0
\end{equation}
can be written as a product of involutions, i.e. $L_c = L_{12}
\circ L_{01} \circ L_{03} \circ L_{04} \circ \dots \circ L_{0n}$
(for $n \ge 3$) \footnote{The three-dimensional case is $L_c =
L_{12} \circ L_{01}$.}, where the involution $L_{ij}$ is defined
as $v_k'=v_k$ ($k \neq i,j$) together with $v_i'=v_j$ and
$v_j'=v_i$ ($i<j$).

The mapping (\ref{L1}) can be written as $L_1=L_z \circ L_c = L_z
\circ L_{12} \circ L_{01} \circ L_{03}$, where
\begin{equation}
  w'=w\, , \; \; x'=x \, , \; \;
  y'=y \, , \; \;
  z'=-z+\frac{\beta(w-x+y)^2+\xi}{\beta(w-x+y)+\gamma}
\end{equation}
and $w=v_0$, $x=v_1$, $y=v_2$ and $z=v_3$.

The mapping (\ref{L4sym}) can be written as $L_2=L_z \circ L_y
\circ L_{02} \circ L_{13}$, where $L_z$ is defined as
\begin{equation}
  w'=w\, , \; \; x'=x \, , \; \;
  y'=y \, , \; \;
  z'=-z-\frac{\beta(w+x+y)^2+\epsilon(w+x+y)+\xi_1}{\beta(w+x+y)+\gamma_1},
\end{equation}
$L_y$ is defined as
\begin{equation}
  w'=w\, , \; \; x'=x \, , \; \;
  y'=-y-\frac{\beta(w+x+z)^2+\epsilon(w+x+y)+\xi_0}{\beta(w+x+z)+\gamma_0} \, , \; \;
  z'=z
\end{equation}
and $w=v_0$, $x=v_1$, $y=v_2$ and $z=v_3$.

The mappings given in the papers \cite{ai} and \cite{ai3d} can be
written as a product of involutions or as an involution composed
with a cyclic shift. For the latter, the above discussion shows
that these mappings can be written as a product of involutions.
Finally, for the mapping given in \cite{ai3d}, i.e.
\begin{equation}
  x'=y \, , \; \;
  y'=z \, , \; \;
  z'=x+\frac{A(y-z)}{Byz+C}
\end{equation}
we have $L_3 = L_z \circ L_c \circ L_{-} = L_z \circ L_{12} \circ
L_{01} \circ L_{-} $, where $L_z$ is given by
\begin{equation}
  x'=x \, , \; \;
  y'=y \, , \; \;
  z'=-z+\frac{A(x-y)}{Bxy+C}\, ,
\end{equation}
$L_{-}$ is given by
\begin{equation}
  x'=-x \, , \; \;
  y'=y \, , \; \;
  z'=z
\end{equation}
and $x=v_0$, $y=v_1$ and $z=v_2$. We note that $L_{-}$ is an
involution.

All of the above mappings are invertible since they are a product
of involutions, i.e. $L = L_0 \circ L_1 \circ \dots \circ L_{m-1}
\circ L_m$ (where the $L_i$ are involutions). Their inverse is
$L^{-1} = L_m \circ L_{m-1} \circ \dots \circ L_1 \circ L_0$.

\end{document}